\documentclass[a4paper,12pt]{article}
\usepackage{latexsym}
\usepackage{epsfig}

\usepackage{amsmath}
\usepackage{amsfonts}
\usepackage{amssymb}
\usepackage{graphicx}

\newcommand{\bi}{\begin{itemize}}
\newcommand{\ei}{\end{itemize}}
\newcommand{\beq}{\begin{equation}}
\newcommand{\eeq}{\end{equation}}
\newcommand{\bea}{\begin{eqnarray}}
\newcommand{\eea}{\end{eqnarray}}

\def\r{\rangle}

\def\N{N}

\def\s{\xi}

\def\zetar{\zeta_{\rm r}}
\def\zetari{\zeta_{{\rm r}0}}
\def\zetarad{\zeta_{\rm r}}
\def\i{{\rm inf}}

\def\k{{\vec k}}

\def\tOm{{\tilde\Omega}}
\def\curv{\sigma}
\def\f{f_c}
\def\r{r}

\def\cf{\sigma}
\def\cs{\chi}
\def\xrf{x_{r1}}
\def\xcf{x_{c1}}
\def\fcf{f_{c1}}
\def\xcc{x_{\cs 1}}
\def\fcc{f_{\cs 1}}
\def\xrs{x_{r2}}
\def\xcs{x_{c2}}
\def\fcs{f_{c2}}

\def\O{{\cal O}}
\def\veps{\epsilon}
\def\vepsf{\varepsilon_f}

\def\rratio{\tilde r}

\def\A{z_\sigma}
\def\B{z_\chi}
\def\C{z_{\sigma\chi}}
\def\D{z_{\sigma\sigma}} 
\def\E{z_{\chi\chi}}
\def\F{s_\sigma}
\def\G{s_\chi}
\def\H{s_{\sigma\chi}}
\def\I{s_{\sigma\sigma}}
\def\J{s_{\chi\chi}}
\def\Lc{s_c}
\def\Kc{z_c}
\def\betaL{\Lambda}
\def\LambdaXi{\Xi}
\def\fNLt{\tilde f_{NL}}

\begin{document}

\title{General treatment of isocurvature perturbations and non-Gaussianities}

\author{David Langlois$^{\, a}$ and  Angela Lepidi$^{\, a,b,c}$
\vspace{0.2cm}
\\
{\small 
$^a$ APC (CNRS-Universit\'e Paris 7), }\\
\vspace{0.2cm}
{\small 10, 
rue Alice Domon et L\'eonie Duquet, 75205 Paris Cedex 13, France}\\
{\small
$^b$ Istituto Nazionale di Fisica Nucleare, Sezione di Ferrara, I-44100 Ferrara, Italy}\\
{\small
$^c$ Dipartimento di Fisica, Universit\`a degli Studi di Ferrara, I-44100 Ferrara, Italy}\\
}

\vspace{1cm}

\maketitle

\begin{abstract}
We present a general formalism that provides a systematic computation of the linear and non-linear perturbations for an arbitrary number of cosmological fluids in the early Universe going through various transitions, in particular the decay of some species (such as a curvaton or a modulus). Using this formalism, we revisit the question of isocurvature non-Gaussianities in the mixed inflaton-curvaton scenario and show that one can obtain significant non-Gaussianities dominated by the isocurvature mode while satisfying the present constraints on the isocurvature contribution in the observed power spectrum. We also study two-curvaton scenarios, taking into account the production of dark matter, and investigate in which cases significant non-Gaussianities can be produced. 
\end{abstract}

\maketitle


\section{Introduction}
In many occasions, cosmology has been and still is an invaluable means to constrain particle physics models. These constraints can arise by using information from homogeneous cosmology, such as the constraints on the light degrees of freedom at nucleosynthesis.  With the discovery of the Cosmic Microwave Background (CMB) fluctuations, new constraints arise  from the observed power spectrum of linear perturbations. Even more recently, the upper bounds on primordial non-Gaussianities have started to be  used to constrain early Universe scenarios.

Although the simplest early Universe models are based on inflationary models with a single scalar field, many models involve additional scalar fields, which can play a dynamical role during inflation or simply be spectactor fields (see e.g. \cite{Langlois:2010xc} for introductory lectures). The existence of several degrees of freedom opens up the possibility of isocurvature perturbations, i.e. perturbations in the particle  density ratio between two fluids, for example cold dark matter  (CDM) isocurvature perturbations (between CDM and radiation) or baryon isocurvature perturbations (between baryons and radiation). Since primordial isocurvature perturbations leave distinctive features in the CMB anisotropies, they can be in principle disentangled from the usual adiabatic mode. The present upper bound  on the isocurvature contribution to the power spectrum provides a stringent constraint.

This is the case for the curvaton scenario~\cite{curvaton} where large residual isocurvature perturbations (for CDM or baryons) can be generated, depending on how and when CDM or baryons are produced~\cite{Lyth:2002my, Lyth:2003ip} (see also \cite{Lemoine:2006sc,Lemoine:2008qj} for more detailed scenarios). The same constraints apply to moduli that are light during inflation, and thus acquire super-Hubble fluctuations, as discussed recently in  \cite{Lemoine:2009is}.  

Another potentially useful information on primordial perturbations is the amplitude and shape of their non-Gaussianity. So far, the current CMB data seem to favour a non-zero amount of so-called local non-Gaussianity~\cite{Komatsu:2010fb}, but Planck data will be needed to confirm or infirm this trend. Several models can generate local non-Gaussianity (see e.g. \cite{Wands:2010af} for a recent review): multiple field inflation (during inflation or at the end of inflation: see e.g. \cite{Byrnes:2010em}), modulated reheating~\cite{Zaldarriaga:2003my,Vernizzi:2003vs}, curvaton, modulated trapping~\cite{Langlois:2009jp}, etc. 
 It is thus interesting to combine the constraints on isocurvature modes and non-Gaussianity to explore the early Universe physics, as has been done recently in various scenarios  \cite{beltran,Kawasaki:2008sn,Kawasaki:2008jy,Langlois:2008vk,Kawasaki:2008pa,Hikage:2008sk, Kawakami:2009iu,Takahashi:2009cx}. 

The purpose of the present work is to give a unified treatment of linear and nonlinear perturbations, which enables to compute their  evolution through one or several cosmological transitions, such as the decay of some particle species. 
Our treatment  takes into account the various decay products and their branching ratio. 
Our formalism can thus be  applied to a large class of early Universe scenarios, in order to compute automatically their predictions for  adiabatic and isocurvature perturbations, and their non-Gaussianities. As input, one simply needs
 parameters that depend on the homogeneous evolution. This thus provides a simple way to confront 
 an early Universe scenario, and  its underlying particle physics model,  with the present and future cosmological data.

As applications to our general formalism, we consider two specific examples. The first example is a more refined treatment of the isocurvature perturbations and their non-Gaussianity in the mixed curvaton-inflation scenario~\cite{Langlois:2004nn,Ferrer:2004nv,Lazarides:2004we}. 
The second example deals with a multiple-curvaton scenario \cite{Hamaguchi:2003dc,Choi:2007fya,Assadullahi:2007uw,Huang:2008rj}. In both examples, we generalize the results that have been obtained in previous works, allowing the curvaton to decay into several species. 

This paper is organized as follows. In Section 2, we introduce the non-linear curvature and isocurvature perturbations. 
Section 3 is devoted to the general treatment of a cosmological transition, such as the decay of some particle species. 
In Section 4, we focus on the first application, namely the mixed curvaton-inflaton scenario with a single curvaton. 
In Section 5, we consider scenarios with two curvatons. We conclude in the final Section.

\section{Non-linear curvature perturbations}
We first introduce  the notion of non-linear curvature perturbation. Several definitions have been proposed, which turn out to be equivalent on large scales, and we will follow here the covariant approach introduced  in
\cite{Langlois:2005ii,Langlois:2005qp}, and reviewed recently in \cite{Langlois:2010vx}. 

For a perfect fluid characterized by the energy density $\rho$, the pressure $P$ and the four-velocity $u^a$, the conservation law $\nabla_a T^{a}_{\ b}=0$ for the energy-momentum tensor,
 $T_{ab}=\left(\rho+P\right) u_a u_b
+P g_{ab}$, 
implies that the covector
\beq
 \zeta_a\equiv
\nabla_a\N-\frac{\dot\N}{\dot\rho}\nabla_a\rho
\label{zeta_a}
\eeq
satisfies the relation
\beq
\label{dot_zeta}
\dot\zeta_a\equiv {\cal L}_u\zeta_a=
-\frac{\Theta}{3(\rho+p)}\left( \nabla_a p -
\frac{\dot p}{\dot \rho} \nabla_a\rho\right) \;,
\eeq
with the definitions 
\beq
 \Theta\equiv\nabla_a u^a, \quad \N\equiv\frac{1}{3}\int d\tau \,
\Theta \;,
\eeq
where $\tau$ is the proper time along the fluid worldlines and a dot denotes a Lie derivative along $u^a$, which is equivalent to an ordinary
derivative for {\it scalar} quantities (e.g. $\dot\rho\equiv u^a\nabla_a\rho$).  $\N$ can be interpreted as  the number of e-folds of the local scale factor  associated with  an observer following the fluid.

The covector $\zeta_a$ can be defined for the global cosmological fluid or for any of the
individual cosmological fluids, as long as they are non-interacting (the case of interacting fluids is discussed in \cite{Langlois:2006iq}). Using
the non-linear conservation equation
\beq
\dot\rho=-3\dot\N(\rho+P)\;,
\eeq
which follows from $u^b\nabla_a T^a_{\ b}=0$,
one can re-express  $\zeta_a$ in the form
\beq
\label{zeta_a2}
\zeta_a=\nabla_a\N+\frac{\nabla_a\rho}{3(\rho+P)} \;.
\eeq
If $w\equiv P/\rho$ is constant,
the above covector  is a total gradient and can be written as
\beq
\zeta_a=\nabla_a\left[\N+\frac{1}{3(1+w)}\ln \rho\right]\, .
\label{zeta_a_w}
\eeq

On scales larger than the Hubble radius, the above definitions are
equivalent to the non-linear curvature perturbation on uniform
density hypersurfaces as defined in
\cite{Lyth:2004gb},
 \beq \zeta = \delta \N -
 \int_{\bar \rho}^{\rho} H \frac{d \tilde \rho}{\dot{ \tilde \rho}}
 =
 \delta \N + \frac13\int_{\bar \rho}^{\rho} \frac{d
 \tilde{\rho}}{(1+w)\tilde{ \rho}}\;, \label{zeta}
 \eeq
where  $H=\dot a /a$ is the Hubble parameter.

It will be useful to distinguish  the non-linear curvature
perturbation $\zeta$ of the total fluid, from the individual non-linear
perturbation $\zeta_A$ that describes the cosmological fluid $A$ (with $w_A\equiv P_A/\rho_A=0$ for a pressureless fluid or 
$w_A=1/3$ for a relativistic fluid), 
 defined by
 \beq
 \label{defzeta}
  \zeta_{_A} = \delta \N +
\frac{1}{3(1+w_{_A})} \ln\left( \frac{\rho_{_A}}{\bar
    \rho_{_A}} \right)\; ,
 \eeq
where a bar denotes a homogeneous quantity.

Inverting this relation yields the expression of the inhomogeneous energy density as a function of the background energy density and of the curvature perturbation $\zeta_{_A}$,
\beq
\label{zeta_deltaN}
\rho_{_A}={\bar
    \rho_{_A}} e^{3(1+w_{_A}) ( \zeta_{_A} -\delta N)}\, ,
\eeq
which we will use many times in the following. 

The non-linear isocurvature (or entropy) perturbation between two fluids $A$ and $B$ is   defined  by 
\beq
S_{A,B}\equiv 3(\zeta_A-\zeta_B).
\eeq
In the following, we will always define the isocurvature perturbations with respect to the radiation fluid, so that our definition for the isocurvature perturbation of the fluid $A$ will be 
\beq
\label{iso_pert}
S_A\equiv 3(\zeta_A-\zeta_r),
\eeq
where $\zeta_r$ is the uniform-density curvature perturbation of the radiation fluid.

\section{Decay}
Let us now consider a cosmological transition associated with the decay of some species of particles, which we will call  $\curv$. 

In the sudden decay approximation, the decay  takes place on the  hypersurface characterized by the condition
\beq
H_{\rm d}=\Gamma_\curv\, .
\eeq
Therefore, since $H$ depends only on  the {\it total} energy density, the decay hypersurface is a hypersurface of uniform total energy density, with $\delta N_{\rm d}=\zeta$, where $\zeta$ is the global curvature perturbation. Using 
(\ref{zeta_deltaN}), the equality between the total energy densities, respectively before and after the decay, thus reads
\beq
\label{bilan_decay}
\sum_A\bar{\rho}_{A-}e^{3(1+w_A)(\zeta_{A-}-\zeta)}=\bar{\rho}_{\rm decay}=\sum_B\bar{\rho}_{B+}e^{3(1+w_B)(\zeta_{B+}-\zeta)},
\eeq
where the subscripts 
$-$  and $+$ label  quantities defined, respectively,  {\it before} and {\it after} the transition. 

\subsection{Before the decay}
The first equality in (\ref{bilan_decay}) leads to
\beq
\label{zeta_non_linear}
\sum_A \Omega_{A}e^{3(1+w_A)(\zeta_{A-}-\zeta)}=1,
\eeq
where we have defined  
$\Omega_A\equiv {\bar\rho_{A-}}/\bar\rho_{\rm decay}$
(to avoid confusion, the $\Omega_A$ are always defined just {\it before} the decay). 
The above relation  determines $\zeta$ as a function of the $\zeta_{A-}$. 

 At linear order, this gives
 \beq
 \label{zeta_decay}
 \zeta=
 \frac{1}{\tOm}  \sum_A\tOm_A\, \zeta_{A-}\qquad ({\rm first}\  {\rm order})
 \eeq
 with the notation
 \beq
\tOm_A\equiv (1+w_A) \Omega_A, \qquad \tOm\equiv\sum_A\tOm_A\, .
\eeq
Expanding (\ref{zeta_non_linear}) 
up to second order, one finds
\beq
\zeta=\frac{1}{\tOm}  \sum_A\tOm_A\, \left[\zeta_{A-}+\frac32 (1+w_A)\left(\zeta_{A-}-\zeta\right)^2\right] \qquad ({\rm second}\  {\rm order})
\eeq
where, on the right hand side, $\zeta$ is to be replaced by its first order expression (\ref{zeta_decay}).

\subsection{After the decay}
We now consider the outcome of the decay. In general, the species $\sigma$ decays 
into various  species $A$, with respective decay widths $\Gamma_{A\curv}$.
Defining the relative branching ratios
\beq
\gamma_{A\curv}\equiv\frac{\Gamma_{A\curv}}{\Gamma_\curv}, \quad \Gamma_\curv\equiv \sum_A \Gamma_{A\curv}\, ,
\eeq
one can write the energy density of the fluid $A$ after the decay in terms of the energy densities of $A$ and of $\curv$ as 
\beq
\rho_{A+}=\rho_{A-}+\gamma_{A\curv}\rho_{\curv -}\, .
\eeq
Using (\ref{zeta_deltaN}), one can rewrite this nonlinear equation in terms of the curvature perturbations $\zeta_{A+}$, $\zeta_{A-}$ and $\zeta_{\curv-}$, which yields
\beq
\label{master_eq}
e^{3(1+w_A)(\zeta_{A+}-\zeta)}=\frac{\bar{\rho}_{A-}e^{3(1+w_A)(\zeta_{A-}-\zeta)}+\gamma_{A\sigma}\bar{\rho}_{\sigma -}e^{3(1+w_\sigma)(\zeta_{\sigma -}-\zeta)}}{\bar{\rho}_{A-}  +\gamma_{A\sigma}\bar{\rho}_{\sigma -}} \, .
\eeq
This expression thus gives   $\zeta_{A+}$ as a function of $\zeta_{A-}$, $\zeta_{\sigma}$ and of the global $\zeta$. Substituting  $\zeta$ in terms of $\zeta_{\sigma}$ and of all the $\zeta_{B-}$, one finally obtains $\zeta_{A+}$ as a function of all the $\zeta_{B-}$.

Following this procedure, one finds that the linear curvature perturbation for any given fluid $A$ is given by
\beq
\label{decay_linear}
\zeta_{A+}=\sum_B T_{A}^{\ B}\zeta_{B-}\qquad ({\rm first}\  {\rm order})
\eeq
with
\begin{eqnarray}
T_{A}^{\ A}&=& 1-f_A+ f_A\frac{(w_A-w_\curv )\tOm_A}{(1+w_A)\tOm}
\label{T1}
\\
T_{A}^{\ \sigma}&=& f_A\, \frac{1+w_\curv}{1+w_A}+f_A\frac{(w_A-w_\curv )\tOm_\curv}{(1+w_A)\tOm}
\label{T2}
\\
T_{A}^{\ C}&=& f_A\frac{(w_A-w_\curv )\tOm_C}{(1+w_A)\tOm}, \qquad C\neq A, \sigma\, .
\label{T3}
\end{eqnarray}
In the above expressions, we have introduced the parameter
\beq
\label{f_A}
f_A\equiv \frac{\gamma_{A\sigma}\Omega_{\sigma}}{\Omega_{A}+\gamma_{A\sigma}\Omega_{\sigma}}\, ,
\eeq
which represents the fraction of the fluid $A$ that has been created by the decay. If $A$ does not belong to the decay products of $\sigma$, then $f_A=0$. The opposite limit, $f_A=1$, occurs when all the fluid $A$ is produced by the decay. For the intermediate values of $f_A$, part of $A$ is produced by the decay while the other part is preexistent.

In the following, we will assume that 
the decaying species behaves like non-relativistic matter (this is the case for a curvaton or modulus field that  oscillates in a quadratic potential)
 and we will thus always use
$ w_\sigma = 0$.

From the above expressions (\ref{T1}-\ref{T3}), it is straightforward to check that 
\beq
\label{sum1}
\sum_B T_{A}^{\ B}=1.
\eeq
The post-decay perturbation $\zeta_{A+}$ can thus be seen as the barycenter of the pre-decay perturbations $\zeta_{B-}$ with the weights $T_{A}^{\ B}$ (all these coefficients satisfy $0\leq T_{A}^{\ B}\leq 1$ 
for $w_\curv=0$). Note that 
if the fluid $A$ is not produced in the decay (i.e. $f_A=0$), then the transfer coefficients are trivial: $T_{A}^{\ B}=\delta_{A}^{\ B}$. 

Since it is  convenient to use the same range of species indices  {\it before and after} the transition, we also
introduce the coefficients 
$T_{\sigma}^{\ B}=0$, which imply that $\zeta_{\sigma+}=0$. This convention will be especially useful when one needs to combine several transitions, as we will discuss soon.

At second order, expanding (\ref{master_eq}) and substituting the first order expression (\ref{zeta_decay}) for $\zeta$, one obtains
\beq
\label{zeta_2ndorder}
\zeta_{A+}=\sum_BT_A^{\ B}\zeta_{B-}+\sum_{B, C}U_A^{BC}\zeta_{B-}\zeta_{C-}, 
\qquad ({\rm second}\  {\rm order})
\eeq
with 
\begin{eqnarray}
U_A^{BC}&\equiv&\frac32 \left[T_{AB}(1+w_B)\delta_{BC}+2\frac{\tOm_C}{\tOm}(w_A-w_B)T_{AB}
-(1+w_A) T_{AB} T_{AC}
\right.
\cr
&&\left. \quad
-\frac{\tOm_B \tOm_C}{\tOm^2}\left(1+w_A-\sum_D T_{AD}(1+w_D)\right)\right].
\label{U_ABC}
\end{eqnarray}

The  change of  the various  isocurvature perturbations, defined in (\ref{iso_pert}),   can also be determined by using the above expressions. 
In particular, at  linear order, one finds, using  the property (\ref{sum1}),  the simple expression 
\beq
\label{S_barycenter}
S_{A+}=\sum_B \left(T_A^{\ B}-T_r^{\ B}\right) S_{B-}\, 
\qquad ({\rm first}\  {\rm order}).
\eeq

\subsection{Several transitions}
If the early Universe scenario involves several cosmological transitions, for example several particle decays,  one can use the above expressions successively to determine the final ``primordial'' perturbations, i.e. the initial conditions at the onset of the standard cosmological era. 

For linear perturbations, the expression of the final perturbations as a function of the initial ones, is simply given by 
\beq
\zeta^{(f)}_A=\sum_B T_A^{\ B} \zeta^{(i)}_B, \qquad T=\prod_k T_{[k]}
\eeq
where $T$ is the matricial product of all transfer matrices $T_{[k]}$, which describe the successive transitions. 

The cosmological transitions can result from the decay of some particle species but they can be of other types.
For example, if CDM consists of  WIMPs (Weakly Interacting Massive Particles),  the freeze-out  can be treated as a cosmological transition.
If radiation is the dominant species at freeze-out, then $\zeta_{c+}=\zeta_r$. But, 
if other species are significant in the energy budget of the universe at the time of freeze-out, any entropy perturbation between the extra species and radiation will modify the above relation. The presence of a pressureless component, like a curvaton, leads to \cite{Lyth:2003ip}
\beq
\label{freeze_out}
\zeta_{c+}=\zeta_{r-}+\frac{(\alpha_f -3)\Omega_{\sigma}}{2(\alpha_f-2)+\Omega_{\sigma}}\left(\zeta_{\sigma -}-\zeta_{r-}\right),\qquad \alpha_f\equiv \frac{m_c}{T_f}+\frac32
\eeq
at linear order, while the other $\zeta_A$ remain unchanged. 
The symbol ``$\sigma$" denotes here the
 conglomerate of all pressureless matter at the time of freeze-out, except of course  the CDM  species that is freezing out.

\section{Scenario with a single curvaton}
\def\xr{x_r}
\def\xc{x_c}
\def\fc{f_c}
Let us now apply our formalism to a simple  scenario with only three initial species: radiation ($r$), CDM ($c$) and a curvaton ($\curv$), considered in e.g. \cite{Gupta:2003jc}. After the decay of the curvaton, the radiation and CDM perturbations remain unchanged and provide the initial conditions for the perturbations at the onset of the standard cosmological phase (let us say around $T\sim 1$ MeV). 

\subsection{Perturbations after the decay}
\subsubsection{Linear order}
According to the expressions (\ref{T1}-\ref{T3}), the linear transfer matrix $T_{AB}$ is given  in this case by  
\beq
\label{T}
T= 
	\left( \begin{array}{ccccc}
	1-\xr && \xc && \xr-\xc  \\
	0 && 1-\fc && \fc  \\
	0  && 0 && 0 
	\end{array} \right), 
	\quad x_r\equiv \frac{f_r}{\tOm}, \quad x_c\equiv \frac14\Omega_c\,  x_r
\eeq
where the order  of the species is $(r,c,\curv)$. This means that 
the linear curvature perturbations for  radiation and for CDM, after the 
 curvaton decay, are given respectively by 
\beq
\zeta_{r+}= (1-x_r)\, \zeta_{r-}+x_c\,  \zeta_{c-}+(x_r-x_c)\, \zeta_{\sigma-}
\eeq
and
\beq
\zeta_{c+}=(1-f_c)\, \zeta_{c-}+f_c \, \zeta_{\curv-}.
\eeq
The entropy perturbation after the decay is thus
\beq
\frac13 S_{c+}\equiv \zeta_{c+}-\zeta_{r+}=(1-f_c-x_c)\zeta_{c-}+(x_r-1)\zeta_{r-}+(f_c+x_c-x_r)\zeta_{\curv-}\,,
\eeq
which can also be expressed directly  in terms of the pre-decay entropy perturbations, following (\ref{S_barycenter}), 
\beq
S_{c+}=(1-f_c-x_c) \, S_{c-}+(f_c+x_c-x_r)\, S_{\curv-}\, .
\eeq
Note that, if many CDM particles are created by the decay of the curvaton, a significant fraction of them could annihilate, leading  to an effective suppression of the final isocurvature perturbation. This effect has been studied in \cite{Lemoine:2006sc} and can easily be incorporated in our formalism.

In practice, we will need the above expressions only  in the limit $x_c=0$ since $\Omega_c$ is usually negligible when the decay occurs. 
The coefficient $x_r$, which we will shorten into $\r$ from now on, can then be expressed as
\beq
\r\equiv x_r= \frac{f_r}{\Omega_\curv}\left(\frac{ 3\Omega_\curv}{4-\Omega_\curv}\right)\equiv \s \, {\tilde r}.
\eeq
The  first factor,
\beq
\s \equiv 
\frac{f_r}{\Omega_\curv}=
 \frac{\gamma_{r \, \curv}}{1-(1- \gamma_{r \, \curv}) \Omega_\curv}
\eeq
can be interpreted as the transfer efficiency between the curvaton and radiation. Its maximal value, $\s=1$, is reached when all the energy stored in the curvaton is converted into radiation, i.e. when $\gamma_{r \, \curv}=1$, as usually assumed in most works on the curvaton.  However, if a fraction of the curvaton energy goes into species other than radiation, then the transfer efficiency $\s$ is reduced.
The second factor, 
\beq
{\tilde r}\equiv\frac{
  3\, \Omega_\curv}{4-\Omega_\curv},
\eeq
  is the familiar coefficient that appears in the literature on the curvaton, which coincides with our $r$ only if  $\s=1$.

\subsubsection{Second order}
The expressions for the curvature perturbations up to second order are obtained from the general expression 
(\ref{zeta_2ndorder}-\ref{U_ABC}), using the transfer matrix (\ref{T}). The expression for  CDM is relatively simple:
\beq
\label{iso_mixed}
\zeta_{c+}=(1-f_c)\zeta_{c-}+f_c \zeta_{\curv-}+\frac32 f_c (1-f_c)\left(\zeta_{c-}-\zeta_{\curv-}\right)^2.
\eeq
The expression for radiation is much more complicated in general, but in the limit  $x_c=0$, which is of interest to us, the radiation perturbation reduces to 
\beq
\label{curv_mixed2}
\zeta_{r+}= \zeta_{r-}+\frac{\r}{3} \, S_{\curv-}+\frac{\r}{18} 
\left[3
-4r 
+\frac{2r}{\xi} - \frac{r^2}{\xi^2} 
\right] S_{\curv-}^2 \,. 
\eeq
In the limit $ \gamma_{r\curv}=1$, i.e. $\s=1$, one recovers the usual expression. 

Note that, although $\Omega_c$ is assumed to be very small, it cannot be neglected in the  expression for $\f$  [see (\ref{f_A})] because $\gamma_{c\curv}$ or $\Omega_\sigma$ can be very small, and $\f$ can thus take any value between $0$ and $1$.

\subsection{Initial curvaton perturbation}
We now need to relate the perturbation of the curvaton fluid with the fluctuations of the curvaton scalar field during inflation. For simplicity, we assume here that the potential of the curvaton is quadratic.

Before its decay, the oscillating curvaton (with mass $m\gg H$) is  described by a pressureless, non-interacting fluid with
energy density
 \begin{equation}
\rho_\curv = m^2 \curv^2 \;,
 \end{equation}
where $\curv$ is the rms amplitude of the curvaton field. Making use of Eq.~(\ref{zeta_deltaN}), the inhomogeneous 
energy density of the curvaton can be  reexpressed as 
 \begin{equation}
  \rho_\curv = \bar\rho_\curv e^{3(\zeta_\curv-\delta N)} \label{rho_curvaton}\;.
 \end{equation}
In the post-inflation era where the curvaton is still subdominant, the spatially flat
hypersurfaces are characterized by $\delta N=\zeta_r$ (since CDM is also subdominant). On such a hypersurface, the
curvaton energy density can be written as
 \begin{equation}
 \bar\rho_\curv e^{3(\zeta_\curv-\zeta_r)}=\bar\rho_\curv e^{S_\curv}
 =
 m^2 \left( \bar\curv+\delta\curv \right)^2 \,. \label{rhorhobarcurv} 
 \end{equation}

Expanding this expression up to  second
order,  and using the conservation of $\delta\curv/\curv$ in a quadratic potential, we obtain
 \beq
S_\curv= 2\frac{\delta\curv_*}{\bar\curv_*}
-\left(\frac{\delta\curv_*}{\bar\curv_*} \right)^2 \,,
 \label{S_dchi}
 \end{equation}
where the initial curvaton field perturbation, $\delta\curv_*$, is assumed to be
Gaussian, as would be expected for a weakly coupled field.
The curvaton entropy perturbation (\ref{S_dchi}) thus contains a linear part $\hat S$
which is Gaussian and a second order part which is quadratic in
$\hat S$:
 \beq
 \label{S_G}
S_\curv=\hat S-\frac14 \hat S^2\, ,
  \quad {\rm with}\quad
 \hat S \equiv 2\frac{\delta\curv_*}{\bar\curv_*}\, 
 \eeq
 where the power spectrum of $\hat S$, generated during inflation,  is given by
 \beq
 \langle \hat S(\vec{k}) \hat S (\vec{k}')\rangle=
 (2\pi)^3 \, \frac{2\pi^2}{k^3}\,{\cal P}_{\hat S}(k) \, \delta(\vec k + \vec k') , 
 \quad 
 {\cal P}_{\hat S}(k)=\frac{4}{\curv_*^2}\left(\frac{H_*}{2\pi }\right)^2
  \;.
 \eeq
 The subscript $*$ means that the quantity is evaluated at the time when the corresponding scale crossed out the Hubble radius during inflation.

\subsection{Primordial adiabatic and isocurvature perturbations}

For simplicity, we now assume that the post-inflation perturbations for dark matter and radiation, i.e. before the curvaton decay, are the same and depend only on the inflaton fluctuations,
\beq
\label{init_inf}
\zeta_{c-}=\zeta_{r-}=\zeta_\i\,,
\eeq
so that there are only two independent degrees of freedom from the inflationary epoch, $\zeta_\i$ and $\hat S$.

Substituting  (\ref{S_G})  and (\ref{init_inf}) into (\ref{curv_mixed2}) and (\ref{iso_mixed}) yields
\beq
\label{zetarad}
\zetarad =  \zeta_\i+\frac{r}{3} \hat S+ \frac{r}{36}\left[3 -8\r 
+\frac{4r}{\xi} -2\frac{r^2}{\xi^2} \right]
\hat S^2\eeq
and
\begin{eqnarray}
\label{S_c}
S_c=(\f-r) \hat S+\frac{1}{12}\left[3 \f (1-2\f)-r\left(3 -8\r 
+\frac{4r}{\xi} -2\frac{\r^2}{\xi^2} \right)\right] \hat S^2,
\end{eqnarray}
In the limit $ \gamma_{r\curv}=1$, i.e. $\s=1$,  one recovers the well-known expression for $\zeta_r$.

Considering only the linear part of (\ref{zetarad}), one finds that
the power spectrum for the primordial adiabatic perturbation
$\zetarad$ can be expressed as
 \beq
 \label{finalzetar}
  {\cal P}_{\zetarad}={\cal P}_{\zeta_\i}+\frac{r^2}{9}{\cal P}_{\hat S} \equiv (1+\lambda){\cal P}_{\zeta_\i}\equiv
  \,  \Xi^{-1}\,  \frac{r^2}{9}{\cal P}_{\hat S}
 \eeq
where $\lambda$ is defined as the ratio between the curvaton and inflaton contributions
and 
$\Xi = (1+\lambda^{-1})^{-1}$ as the ratio between the curvaton contribution and the total curvature power spectrum.
The limit $\lambda\gg 1$, or $\Xi\simeq 1$, corresponds to  the standard curvaton scenario, where
the inflaton perturbation is ignored. 
The cases where the inflaton contribution is not negligible correspond to the mixed inflaton-curvaton scenario~\cite{Langlois:2004nn}. The curvaton contribution is subdominant when $\lambda\ll 1$, i.e. $\Xi\ll 1$. 

Let us now turn to the primordial isocurvature perturbation. As can be read from the linear part of (\ref{S_c}), its power spectrum is given by 
\beq
{\cal P}_{S_c}=(\f-r)^2 {\cal P}_{\hat S}\, .
\eeq
and the correlation between adiabatic and isocurvature fluctuations is 
\beq 
\label{corr_ad_is}
{\cal C}
\equiv \frac{{\cal P}_{S_c, \zeta_r}}{\sqrt{{\cal P}_{S_c} {\cal P}_{\zeta_r}}}= \vepsf
\, \Xi^{1/2}\,, \qquad \vepsf \equiv {\rm sgn}(f_c-r)\, .
\eeq
In the pure curvaton limit ($\Xi\simeq 1$), adiabatic and isocurvature perturbations are either fully correlated, if $\vepsf>0$, or fully anti-correlated, if $\vepsf<0$.
In the opposite limit ($\Xi\ll 1$), the correlation vanishes. 
For intermediate values of  $\Xi$, the correlation is only partial, as can be also obtained  in multifield inflation
\cite{Langlois:1999dw}.

As combined adiabatic and isocurvature perturbations lead to a distortion of the acoustic peaks, which depends on their correlation~\cite{Langlois:2000ar}, it is in principle possible to distinguish, in the observed fluctuations,  the adiabatic and isocurvature contributions. So far, there is no detection of any isocurvature component, but only an upper bound on the ratio between isocurvature and adiabatic power spectra, which, in our case, is given by
\beq
\label{alpha}
\alpha\equiv \frac{{\cal P}_{S_c}}{{\cal P}_{{\zeta}_{\rm r}}} = 9 \left(1-\frac{\f}{\r}\right)^2 \, \Xi   \, .
\eeq
The observational constraints on $\alpha$ depend on the correlation. Writing $\alpha\equiv a/(1-a)$ (note that $\alpha\simeq a$ if $\alpha$ is small), the constraints (WMAP+BAO+SN) given in  \cite{Komatsu:2010fb} are 
\beq
a_0<0.064\quad (95 \% {\rm CL}), \qquad a_1< 0.0037 \quad (95 \% {\rm CL})
\eeq
respectively for the uncorrelated case and for the fully correlated case
\footnote{Our notations differ from those of \cite{Komatsu:2010fb}. Our $a$ corresponds to their $\alpha$ and our fully {\sl correlated} limit corresponds to their fully {\sl anti-correlated} limit, because their definition of the correlation has the opposite sign.}.

One sees that the observational constraint $\alpha \ll 1$ can be satisfied in only two cases:
\bi
\item $|\f-r|\ll r$, i.e. a fine-tuning between the two parameters $\f$ and $\r$.  This includes the case   $\f=1$ with  $r\simeq  1$, considered in \cite{Langlois:2008vk}.
\item $\Xi \ll 1$, i.e. the curvaton contribution to the observed power spectrum is very small.
\ei

\subsection{Non-Gaussianities}
\label{sec-non-gauss}
Let us now examine the amplitude of the non-Gaussianities that can be generated in our model. 
We recall that   our  observable quantities $\zeta$ and $S$ are of  the form 
\beq
\zeta=\zeta_\i+z_1\,  \hat S+\frac12 z_2\,  \hat S^2, \qquad S=s_1 \, \hat S+\frac12 s_2\,  \hat S^2,
\eeq
where $\zeta_\i$ and $\hat S$ are two independent Gaussian fields and where the coefficients can be  read explicitly from (\ref{zetarad}) and (\ref{S_c}).

Applying the general results of the Appendix to the present situation, we  can easily compute  for our model the ``reduced" angular bispectrum, which is of direct interest for a comparison with CMB observations and which generalizes the analysis of \cite{Komatsu:2001rj} in the purely adiabatic case. Specializing 
(\ref{reduced_bispectrum_general}) to our case, one finds 
\begin{eqnarray}
\label{b_l}
b_{l_1l_2 l_3}= 3  \sum_{I,J,K}\,  b_{NL}^{I, JK} \int_0^\infty r^2 dr \tilde\beta^I_{(l_1}(r)\beta^{J}_{l_2}(r)\beta^{K}_{l_3)}(r)
\end{eqnarray}
with  
\beq
b_{NL}^{I, JK} \equiv N^I_{(2)} N^J_{(1)} N^K_{(1)},  
\eeq
where $N^\zeta_{(2)}=z_2$, $N^S_{(2)}=s_2$, $N^\zeta_{(1)}=z_1$, $N^S_{(1)}=
s_1$, respectively, and 
\beq
\label{beta}
\tilde\beta^I_{l}(r)\equiv \frac{2}{\pi} \int k^2 dk \,  j_l(kr) g^I_{l}(k) , \quad 
\beta^{I}_{l}(r)\equiv \frac{2}{\pi}\int k^2 dk\,  j_l(kr) g^I_{l}(k) P_{\hat S}(k)\, ,\quad
\eeq
where  the $ g^I_{l}(k)$ denote the adiabatic ($I=\zeta$) and isocurvature ($I=S$) transfer functions.  
Because of the symmetry under exchange of the last two indices, the reduced bispectrum contains six different contributions, whose amplitude is parametrized by the six coefficients $b_{NL}^{I, JK} $.

In order to compare these coefficients with the usual parameter $f_{NL}$  defined in the purely adiabatic case, one must recall that $f_{NL}$ is proportional to the bispectrum of $\zeta$ divided by the square of its power spectrum. By  noting that the $\beta_l^I(r)$ introduced in (\ref{beta}) involve  $P_{\hat S}$, this implies that  the analogs of $f_{NL}$ can be defined  by dividing  the coefficient $b_{NL}^{I, JK}$ by the square  of the ratio 
$P_\zeta/P_{\hat S}=z_1^2\Xi^{-1}$. We thus introduce 
 the parameters
\beq
\fNLt^{I, JK} \equiv \frac65 f_{NL}^{I, JK} \equiv \frac{\Xi^2}{z_1^4} \, b_{NL}^{I, JK} ,
\eeq
explicitly given by the expressions
\begin{eqnarray}
&\fNLt
^{\,\zeta, \zeta\zeta}=\frac{z_2}{z_1^2}\, \Xi^2,  \quad \fNLt^{\zeta, \zeta S}=\frac{s_1 z_2}{z_1^3}\, \Xi^2,  \quad 
\fNLt^{\zeta, S S}=\frac{s_1^2 z_2}{z_1^4}\, \Xi^2, &
\\\cr
&\fNLt^{S, \zeta\zeta} = \frac{s_2}{z_1^2}\, \Xi^2,
\quad \fNLt^{S, \zeta S}= \frac{s_1 s_2}{z_1^3}\, \Xi^2,  \quad 
\fNLt^{S, SS}= \frac{s_1^2 s_2}{z_1^4}\, \Xi^2\, . &
\end{eqnarray}

In the absence of  isocurvature perturbations, the above non-linear parameters vanish 
except the first one, yielding 
  \beq
f_{NL}^{\zeta, \zeta\zeta} =\frac56 \,
\left(\frac{3}{2r} +\frac{2}{\xi}-4 -\frac{r}{\xi^2} \right)\, \Xi^2
\label{f_ad}\;,
 \eeq
 which exactly coincides with the familiar  parameter $f_{NL}$. 
 The amplitude  of the  non-gaussianities is   determined by the three parameters $\r $, $\s$ and $\Xi$ (note  that   one recovers the usual prediction of the pure curvaton scenario for $\s=1$ and $\Xi=1$), which take values between $0$ and $1$. 
The present constraints on  $f_{NL}$, calculated from WMAP data by assuming purely adiabatic perturbations, are  \cite{Komatsu:2010fb}: 
	\begin{eqnarray}
	-10 \leq f_{NL}^{(\mathrm{local})} \leq 74.
	\end{eqnarray}
A sufficiently small $\r$, or even $\s$, leads to a significant non-Gaussianity from the adiabatic component, whereas a small $\Xi$ tends to suppress it. 

If isocurvature modes are present, however,  the five  other terms in the reduced bispectrum (\ref{b_l}) will also contribute in general. Interestingly,  the six functions 
on the right hand side of  (\ref{b_l})  have distinct dependences on the $l_i$, because they involve different combinations of the adiabatic and isocurvature transfer functions. Therefore, this allows in principle to measure, or constrain, independently  the corresponding six non-linear parameters from the CMB data. 
The precise determination of constraints on the $f_{NL}^{I,JK}$ is beyond the scope of the present work, 
but since all the functions multiplying the $b_{NL}^{I,JK}$ in (\ref{b_l}) are of similar amplitude, one can a priori expect  the constraints on the $f_{NL}^{I,JK}$ to be of the same order of magnitude as  those  on $f_{NL}$\footnote{Observational constraints on isocurvature non-Gaussianities are given in  \cite{Hikage:2008sk} , for an isocurvature perturbation of the form $S=S_L+f_{NL}^{(iso)}S_L^2$, 
where $S_L$ is Gaussian. Their non-linear  parameter $f_{NL}^{(iso)}$ is related to ours according to $\fNLt^{S, SS}= 2 f_{NL}^{(iso)}\alpha^2$, $\fNLt^{S, \zeta S}=2f_{NL}^{(iso)} \alpha^{3/2} |{\cal C}|$ and 
$\fNLt^{S, \zeta\zeta} =2 f_{NL}^{(iso)}\alpha\,  {\cal C}^2$, where ${\cal C}$ is the  correlation defined in (\ref{corr_ad_is}). }.

Let us now explore the  amplitude of the non-linear parameters in our model. First of all, let us stress that finding significant non-Gaussianities (typically $f_{NL}\sim 10-100$) requires, in all cases, a small denominator $z_1$, i.e.  $r\ll 1$, which will thus be assumed below. 
Second, it is worth noting that all the coefficients are related via  the two rules
\beq
 f_{NL}^{I, JS}= \frac{s_1}{z_1} f_{NL}^{I, J\zeta}\ (R1), \qquad f_{NL}^{S, IJ}= \frac{s_2}{z_2} f_{NL}^{\zeta, IJ}, \ (R2)\, .
\eeq
Therefore, the hierarchy between the parameters can be deduced from the value of the first order ratio
\beq
\label{ratios_sz}
\frac{s_1}{z_1}=3\left(\frac{\f}{\r}-1\right)=\vepsf \sqrt{\frac{\alpha}{\Xi}}\,, \qquad \vepsf\equiv {\rm sgn}(\f-\r),
\eeq
where we have used (\ref{alpha}), as well as the second order ratio $s_2/z_2$, which is a more complicated expression in general.

We now consider successively the two limits for which the isocurvature bound is satisfied.

\subsubsection{Limit $|\f-\r|\ll \r$, with $\Xi\simeq1$ (pure curvaton scenario)}

In this case,
the isocurvature-adiabatic ratio $\alpha$ must satisfy the observational constraint   $\alpha \simeq a_1 \leq 0.0037$, since we are in the  fully correlated case.
The relevant ratios are given here by 
\beq
\frac{s_1}{z_1}\simeq \vepsf \sqrt{\alpha}, \qquad 
\frac{s_2}{z_2}\simeq 
\frac{\vepsf\sqrt{\alpha} -2\rratio (2-\rratio)}{1+2\, \rratio \left(2-\rratio\right)/3} 
\eeq
where we have taken the limit $\r=\s\,\rratio\ll 1$ (although $\r$ cannot be smaller than $10^{-2}$, to be compatible with observational constraint on $f_{NL}$). If $\r$ is small because $\rratio \ll 1$, then the denominator in the expression for  $s_2/z_2$ reduces to $1$. However, if   $\xi\ll 1$ while 
 $\rratio$ is of order $1$, the full expression for $s_2/z_2$ is needed.

The value of the first ratio  implies that, with respect to $f_{NL}^{\zeta, \zeta\zeta}$, the coefficients $f_{NL}^{\zeta, \zeta S}$ and $f_{NL}^{\zeta, S S}$ are suppressed  with factors 
$\sqrt{\alpha}$ and $\alpha$, respectively, according to (R1).  
Analogously the coefficients  $f_{NL}^{S,\zeta S}$ and $f_{NL}^{S,SS}$ are  suppressed, respectively  with factors 
$\sqrt{\alpha}$ and $\alpha$, with respect to $f_{NL}^{S,\zeta \zeta}$. 
By contrast, using (R2), one sees that $f_{NL}^{S,\zeta \zeta}$ could be of the same order of magnitude as $f_{NL}^{\zeta, \zeta\zeta}$, if $\tilde r\sim 1$, or suppressed if $\tilde r$ is small.

To conclude,  in the pure curvaton scenario, it is possible to satisfy the isocurvature  constraint and to get measurable non-gaussianities only by assuming a  fine-tuning between $\f$ and $\r$ at the percent level. In this situation, only the purely adiabatic parameter is significant,  while the other parameters are suppressed, with increasing powers of $\alpha$.

\subsubsection{Limit $\Xi \ll 1$}

In this limit $\alpha$ must satisfy the  constraint $\alpha\simeq a_0<0.064$ (uncorrelated case). 

In the  regime $\f\ll \r\ll 1$, one finds that both ratios $s_1/z_1$ and $s_2/z_2$ reduce  to (-3),  independently of the value of $\rratio$. 
Therefore,  the relation between the non-linear parameters is simply 
\beq
\label{f_tilde_1c_2}
\fNLt^{\zeta, \zeta\zeta}  \simeq \frac{\alpha^2}{54r}, \quad \fNLt^{I, JK}\simeq (-3)^{I_S}\fNLt^{\zeta, \zeta\zeta} \qquad (\f\ll \r\ll 1)
\eeq
where $I_S$ is the number of  $S$ among the indices $(I, J K)$. 
This is the result obtained in \cite{Langlois:2008vk} for $\f=0$. 
For $\alpha$ close to its present upper bound, one sees that detectable 
 non-Gaussianity   can be generated with  $r \sim 10^{-5}$.

By contrast, in the regime $\f\gg r$, the purely adiabatic coefficient is strongly suppressed since 
\beq
\fNLt^{\zeta, \zeta\zeta} \simeq \frac{\alpha^2\r^3}{54 \f^4}\, .
\eeq
However,  the other coefficients are now  enhanced  with respect to the purely adiabatic coefficient, via   the large factors 
\beq
\frac{s_1}{z_1}\simeq 3 \frac{\f}{\r}, \qquad
\frac{s_2}{z_2}\simeq 3\frac{\f}{\r}(1-2\f)
\, .
\eeq
where, for simplicity, we have assumed $\rratio \ll 1$ (the other possibility $\s\ll 1$ yields a more complicated expression for the second ratio, with a dependence on $\rratio$).
The  dominant term is therefore the  purely isocurvature term
\beq
\fNLt^{S, SS} \simeq \alpha^2\, \frac{1-2 \f}{2\f}
\eeq
If $\f\sim 1$, this purely isocurvature non-Gaussianity, although enhanced with respect to all the other contributions, remains  negligible since it is suppressed by the very small factor 
 $\alpha^2$. This was the conclusion reached in  \cite{Langlois:2008vk} (for  $\f=1$).
  
However, we now see that this suppression can be compensated if $\f$ is smaller than $\alpha^2$. The purely isocurvature parameter and the other ones are then given by
\beq
\fNLt^{S, SS}\simeq  \frac{\alpha^2}{2\f},\quad \fNLt^{I, JK}\simeq \left(\frac{\r}{3\f}\right)^{I_\zeta}\fNLt^{S, SS} \qquad (r\ll \f \ll 1)
\eeq
where $I_\zeta$ is the number of $\zeta$ among the three indices.  One can notice that the amplitude of the purely isocurvature non-Gaussianity does not depend on the parameter $\r$, but only on $\alpha$ and $\f$. For instance, with $\alpha=0.05$ which satisfies the current observational bound, a value $\f=10^{-5}$ yields 
$\fNLt^{S, SS}\sim 100$. In such a scenario, one gets observable non-Gaussianity that  comes essentially from isocurvature modes, even if the latter are subdominant in the power spectrum.

\section{Scenario with two curvatons}
We now apply our formalism to the models where two curvatons are present in the early Universe (see e.g. \cite{Hamaguchi:2003dc,Choi:2007fya,Assadullahi:2007uw}).
 The curvaton $\cf$ will be assumed to decay  first,  followed later by the  curvaton denoted $\cs$.

\subsection{First  order}
At linear order, the decay of the first curvaton can be characterized by the transfer matrix
\beq
T_{[1]}= 
	\left( \begin{array}{ccccccc}
	1-\xrf && \xcf && \xcc && \xrf-\xcf  -\xcc \\
	0 && 1-\fcf && 0 && \fcf  \\
	0 && 0 &&  1-\fcc && \fcc \\
	0  && 0 && 0 && 0
	\end{array} \right),
\label{MT1}	 
\eeq
where the order  of the species is $(r,c,\cs, \cf)$, while the decay of the second curvaton is characterized by the transfer matrix
\beq
T_{[2]}= 
	\left( \begin{array}{ccccccc}
	1-\xrs && \xcs && \xrs-\xcs && 0 \\
	0 && 1-\fcs && \fcs && 0  \\
	0 && 0 &&  0 && 0 \\
	0  && 0 && 0 && 0
	\end{array} \right)\, .
\label{MT2}
\eeq
In the above matrices, the definitions of the parameters are analogous to the definitions  introduced in (\ref{T}), i.e.  $\xrf \equiv f_{r1}/{\tOm_1}$, $\xcf\equiv \Omega_{c1}\,  \xrf /4$, $\xcc\equiv \Omega_{\chi 1}\,  \xrf /4$, etc, and the indices $1$ and $2$ refer respectively to the first and second decays.
We have also allowed the possibility that the first curvaton $\cf$ decays into the second curvaton $\cs$,
hence the presence of the parameter $\fcc$.

The expression of the perturbations for radiation and CDM, after the two transitions, are  expressed in terms of the initial perturbations $\zeta_{B0}$ via the  product of the two transfer matrices given above, i.e. 
\beq
\zeta_A=\sum_B \left(T_{[2]}\cdot T_{[1]}\right)_A^{\ B} \zeta_{B0}.
\eeq

At first order, the radiation curvature  perturbation, after the second curvaton decay, reads
\beq
\zetar=\zetari+
\A \, S_{\cf 0}+ 
\B \, S_{\cs 0}+
\Kc \, S_{c 0},
\eeq
with
\begin{eqnarray}
3\A&=& (1-\xrs)(\xrf-\xcf-\xcc)+\fcf \xcs+\fcc (\xrs-\xcs),
\label{A}
\\
3\B&=&(1-\fcc)(\xrs-\xcs)+(1-\xrs) \xcc,
\label{B}
\\
3\Kc &=& (1-\fcf) \xcs+(1-\xrs) \xcf\, .
\end{eqnarray}
Combining this expression with that of the CDM curvature perturbation, according to (\ref{iso_pert}), we find that the CDM entropy perturbation is given by 
\beq
S_c=\F \, S_{\cf 0}+\G \, S_{\cs 0}+\Lc \, S_{c 0},
\eeq
with
\begin{eqnarray}
\F&=& -3\A +\fcf (1-\fcs)+\fcs \fcc,
\label{F}
\\
\G&=& -3\B +\fcs (1-\fcc),
\label{G}
\\
\Lc&=& -3\Kc + (1-\fcf)(1-\fcs).
\end{eqnarray}
For simplicity, we will restrict ourselves, from now on, to the case where $S_{c0}=0$. 

Defining $\betaL$ as the ratio between the two curvaton power spectra, such that
\beq
P_{S_{\cs0}}\equiv\betaL P_{S_{\cf0}},
\eeq
one easily finds that the ratio between the isocurvature and the adiabatic spectra is given by
\begin{eqnarray}
	\label{ratio_spec_beta}
	\alpha =
	\frac{P_{S_c}}{P_{\zetarad}} =
	\frac{\F^2 + \betaL \G^2}{\A^2 + \betaL \B^2}
	\; \LambdaXi\,, 
	\qquad
	\LambdaXi \equiv \frac{\lambda_\chi+ \lambda_\sigma}{1+\lambda_\chi+\lambda_\sigma}
\end{eqnarray}
where $\lambda_\chi$ and $\lambda_\sigma$ are defined as in (\ref{finalzetar}), i.e. 
\beq
 {\cal P}_{\zetarad}={\cal P}_{\zeta_{r0}}+
 \A^2{\cal P}_{S_{\sigma 0}} +
 \B^2{\cal P}_{S_{\chi 0}} \equiv (1+\lambda_\sigma+\lambda_\chi){\cal P}_{\zeta_{r0}}.
 \eeq
 The correlation between $\zetarad$ and $S_c$ can be expressed as
 \beq
 {\cal C}
 =\frac{\A\F+\betaL \B\G}{\sqrt{(\F^2 + \betaL \G^2)(\A^2 + \betaL \B^2)}}\, \sqrt{\LambdaXi}\, .
 \eeq

The observational constraints  on  $\alpha$  impose that   at least one of the following conditions must be satisfied:
\beq
\LambdaXi
\ll 1
 \quad
{\rm or} 
\quad
\F^2 + \betaL \G^2\ll \A^2 + \betaL \B^2
\, .
\eeq
The first possibility, $\LambdaXi \ll 1$, corresponds to a power spectrum dominated by the 
inflaton, whereas the second possibility requires special cancellations in (\ref{F}-\ref{G}) so that $\F$ and $\G$ are suppressed.

\subsection{Second order}
We now consider the perturbations up to the second order, in order to  compute the non-Gaussianities. 
First, let us decompose  the curvaton entropy perturbations as in (\ref{S_G}), so that 
	\begin{eqnarray}
	S_{\cf 0} = \hat S_{\cf} - \frac{1}{4} \hat S_{\cf}^2
	\hspace{1.5cm}
	S_{\cs 0} = \hat S_{\cs} - \frac{1}{4} \hat S_{\cs}^2,
	\end{eqnarray}
where $\hat S_{\cf}$ and $\hat S_{\cs}$ are two independent Gaussian quantities.
	
The radiation curvature perturbation and the dark matter entropy perturbation after the second decay, up to second order,  are given in our notation by
	\begin{eqnarray}
	\zetar &=& \zetari + 
	\A \hat S_{\,\cf} + 
	\B \hat S_{\,\cs} + \C \hat S_{\,\cf} \hat S_{\,\cs} 
	+ \frac{1}{2} \D \hat S_{\,\cf}^2 + \frac{1}{2} \E \hat S_{\,\cs}^2 \label{zetar}
	\\ \cr
	S_c &=&  \F \hat S_{\,\cf} + \G \hat S_{\,\cs} + \H \hat S_{\,\cf} \hat S_{\,\cs} 
	+ \frac{1}{2} \I \hat S_{\,\cf}^2 + \frac{1}{2} \J \hat S_{\,\cs}^2 \label{Sc}
	\end{eqnarray}
where the coefficients $\A$, $\B$, $\F$ and $\G$ have already been defined in (\ref{A}-\ref{B}) and (\ref{F}-\ref{G}), respectively. 
We do not give explicitly the full expressions for the second order coefficients because they are very lengthy, but they are straightforward to compute by using the general expressions (\ref{zeta_2ndorder}-\ref{U_ABC}) with the transfer matrices (\ref{MT1}-\ref{MT2}).

Let us calculate the reduced bispectrum by using the general expression given in the Appendix. In our model, ignoring the inflaton which does not produce significant non-Gaussianities, the relevant power spectra are independent so that 
	\begin{eqnarray}
	P^{ab}(k)=\left( \begin{array}{ccc}
	1 && 0\\ 0 && \betaL
	\end{array} \right)\, P_{\hat S_{\,\cf} }\,,
	\end{eqnarray}
where we furthermore assume that $\betaL$ is strictly independent of $k$ (this is indeed the case if the masses of both curvatons are negligible with respect to $H$ during inflation). 

As a consequence, the reduced bispectrum  can be reduced to the same expression as that already given in Eq. (\ref{b_l}) with
\beq
\label{beta_2}
\beta^{I}_{l}(r)\equiv \frac{2}{\pi}\int k^2 dk j_l(kr) g^I_{l}(k) P_{\hat S_{\,\cf} }(k)\, 
\eeq
and the six  parameters
	\begin{eqnarray}
	\label{b_coeff_2c}
	b_{NL}^{I,JK} \equiv N_{\cf\cf}^I N_{\cf}^J N_{\cf}^K
	+ \betaL N_{\cf\cs}^I \left(N_{\cf}^J N_{\cs}^K+N_{\cs}^J N_{\cf}^K\right)
	+\betaL^2 N_{\cs\cs}^I N_{\cs}^J N_{\cs}^K\,,
	\end{eqnarray}
where the coefficients $N_{ab}^I$, which  are defined as	in (\ref{X_I}), can  be read off directly from (\ref{zetar}) and (\ref{Sc}).
In complete analogy with the model with one curvaton, to be compared with the standard $f_{NL}$, these coefficients must be divided by the square of the ratio $P_{\zeta}/P_{S_{\cf0}} = (\A^2+\betaL\B^2)/\LambdaXi$, hence:
	\begin{eqnarray}
	\label{f_tilde_2c}
	\fNLt^{I,JK} \equiv \left(\frac{\LambdaXi}{\A^2+\betaL\B^2}\right)^2b_{NL}^{I,JK}.
	\end{eqnarray}
The six non-linearity coefficients are thus given by 
	\begin{eqnarray}
	\label{b_NL_2C_lambda}
	\fNLt^{\zeta, \zeta \zeta} &=& 
	\left(\frac{\LambdaXi}{\A^2+\betaL\B^2}\right)^2
	\left[\D\A^2 + 2 \betaL \C\A\B + \betaL^2 \E\B^2  \right],
	\cr\cr
	\fNLt^{\zeta,\zeta S} &=& \left(\frac{\LambdaXi}{\A^2+\betaL\B^2}\right)^2
	\left[\D\A\F+\betaL\,\C(\A\G+\B\F)+\betaL^2\E\B\G \right],
	\cr\cr
	\fNLt^{\zeta, SS} &=& \left(\frac{\LambdaXi}{\A^2+\betaL\B^2}\right)^2
	\left[\D\F^2+2\betaL\,\C\F\G+\betaL^2\E\G^2 \right],
	\cr\cr
	\fNLt^{S,\zeta\zeta} &=& \left(\frac{\LambdaXi}{\A^2+\betaL\B^2}\right)^2
	\left[\I\A^2 + 2\betaL\H\A\B+\betaL^2\J\B^2 \right],
	\cr\cr
	\fNLt^{S,S\zeta} &=& \left(\frac{\LambdaXi}{\A^2+\betaL\B^2}\right)^2
	\left[\I\A\F + \betaL\H(\F\B+\G\A) + \betaL^2\J\G\B \right],
	\cr\cr
	\fNLt^{S, SS} &=& 
	\left(\frac{\LambdaXi}{\A^2+\betaL\B^2}\right)^2
	\left[\I \F^2 + 2 \betaL \H\F\G + \betaL^2 \J\G^2   \right]\,.	
	\end{eqnarray}
In the following we analyze explicitly some limiting cases.

\subsection{Various limits}

We now explore the parameter space, in order to see whether it is possible to obtain significant non-Gaussianities. 

Let us first mention that we have checked that our results agree with those of \cite{Assadullahi:2007uw} in the limit where the curvatons decay only into radiation  (i.e. $\fcf=\fcs=\fcc=0$), the dark matter abundance is neglected (i.e. $\xcf=\xcs=0$) and the inflaton contribution is ignored (i.e. $\LambdaXi=1$).

\subsubsection{Limit $\betaL \ll 1$}
In this limit where the contributions from the second curvaton are negligible, one finds 
\beq
\label{limit_beta_0}
\alpha\simeq \LambdaXi \, \frac{\F^2}{\A^2}, \qquad \fNLt^{\zeta \zeta \zeta}\simeq \LambdaXi^2 \frac{\D}{\A^2}, 
\eeq
while the other five non-linear coefficients can be deduced from $\fNLt^{\zeta \zeta \zeta}$ according to the relations
\beq
\label{rules_c2_1}
f_{NL}^{I, JS}\simeq \frac{\F}{\A} f_{NL}^{I, J\zeta}\, \qquad f_{NL}^{S, IJ}\simeq \frac{\I}{\D} f_{NL}^{\zeta, IJ} \, .
\eeq

The quantity $\alpha$ is constrained by observations to be small, which  requires either $\LambdaXi\ll 1$ or  $|\F|\ll |\A|$.

\paragraph{First possibility: $\LambdaXi \ll 1$,} while $|\A| \sim |\F|$.

This leads to a suppression of all the non-Gaussianity coefficients by a factor $\LambdaXi^2$. However, the coefficients $\fNLt^{\zeta, JK}$ can still  be significant if  the ratio $\D/\A^2$ can compensate the $\Xi^2$ suppression (similarly for the $f_{NL}^{S,JK}$ if  $\I/\A^2 $ compensates the $\Xi^2$ suppression). 

Let us consider a specific example, with the simplifying  assumptions
	\begin{eqnarray}
	\label{par_ex_2c}
	\xcf=\xcs=\fcc=\xcc=0\,,
	\end{eqnarray}
that is, we neglect the energy fraction of dark matter and assume that the curvaton $\cf$ does not decay into $\cs$ and that $\cs$ is subdominant when $\cf$ decays. 
Under these assumptions, 
$\A= \xrf(1-\xrs)/3$ and we further assume $\fcf \ll\A$ so that $\F\simeq -3\A$.
In the two limits $\xrf = \tilde r_1\xi_1\ll1$ and $(1-\xrs)\ll1$, $\A$ is small and 
the adiabatic non-Gaussianity behaves as
	\begin{eqnarray}
	\label{fNL_ex_2c}
	\fNLt^{\zeta, \zeta \zeta}=
	\frac{1}{1-\xrs} \left[\tilde f_{NL\, 1}^{\zeta, \zeta \zeta} + \frac{\xrs}{1-\xrs}
	\left(\frac32+\xrs \,\tilde f_{NL\, 2}^{\zeta, \zeta \zeta} \right) \right]\,,
	\end{eqnarray}
where $\tilde f_{NL\, 1,2}^{\zeta, \zeta \zeta}$ correspond to  single-curvaton coefficient, equation (\ref{f_ad}), but calculated with the parameters $\xi_{1,2}$ and $x_{r1,r2}$ respectively.

If we assume $\xrs\ll1$, the above equation corresponds to the single-curvaton result   (\ref{f_tilde_1c_2}). 
The other coefficients also follow the relations given  in (\ref{f_tilde_1c_2}), since $\I/\D=-3$ with the assumptions (\ref{par_ex_2c}) and $\fcf \ll 1$, and are thus of comparable magnitude.

\paragraph{Second possibility:} $|\F|\ll |\A|$

When a small $\alpha$ is the consequence of $|\F|\ll |\A|$, one sees from the first relation in (\ref{rules_c2_1}) that all the $f_{NL}^{I, JS}$ are strongly suppressed with respect to  $f_{NL}^{I, J\zeta}$. However, the two coefficients $f_{NL}^{I,\zeta\zeta}$ can still be important if $ |\A|$ is sufficiently small. 
By examining (\ref{A}) and (\ref{F}), one sees that getting $|\F|\ll |\A|\ll 1$ requires some fine-tuning between the coefficients, which we now discuss.

In order to get $|\A|\ll 1$, the first possibility is that the first curvaton is subdominant, i.e. $\xrf=\O(\veps)$, where $\veps$ is some small number (we neglect $x_{c2}$ which must be small because we are deep in the radiation era), which then requires either $\xrs=\O(\veps)$ or
$\fcc=\O(\veps)$. The second possibility is that the second curvaton dominates at decay, i.e. $\xrs=1-\O(\veps)$, which also requires that $\fcc=\O(\veps)$. 
Then, to obtain $|\F|\ll |\A|$, the terms of the right hand side of (\ref{F}), which are of order $\veps$ must compensate each other so that their sum is at most of order $\O(\alpha\,\veps)$, which necessitates some special relation between the $f_A$ and the $x_A$.

If we consider  again the assumptions  (\ref{par_ex_2c}) and neglect $\fcs$, one finds that the fine-tuning condition to get  $|\F|\ll |\A|$ is
	\begin{eqnarray}
	\fcf-\xrf(1-\xrs)\leq {\cal O}(\alpha \epsilon)\,.
	\end{eqnarray}
The adiabatic parameter $\fNLt^{\zeta, \zeta \zeta}$ 
is given in equation (\ref{fNL_ex_2c}), with now $\Xi\sim 1$, and its value is of order $10$ when $\epsilon\sim \xrf(1-\xrs) \sim 0.1$. Since $\I/\D\simeq - 3 +\O(\fcf/\xrf(1-\xrs))$,  we also have $\fNLt^{S,\zeta\zeta}\sim\fNLt^{\zeta, \zeta \zeta}$.

Note that a significant non-Gaussianity generated by   a dominant curvaton ($\xrs=1-\O(\veps)$) has already been pointed out in \cite{Assadullahi:2007uw}, but we see here that satisfying the isocurvature bound requires additional constraints on the branching ratios of the curvatons.  

\subsubsection{Limit $\betaL \gg 1$}
In this limit, one obtains
\beq
\alpha\sim  \LambdaXi \, \frac{\G^2}{\B^2}, \qquad \fNLt^{\zeta \zeta \zeta}\sim \LambdaXi^2\, \frac{\E}{\B^2}, 
\qquad f_{NL}^{I, JS}\simeq \frac{\G}{\B} f_{NL}^{I, J\zeta}\, ,
\qquad f_{NL}^{S, IJ}\simeq \frac{\J}{\E} f_{NL}^{\zeta, IJ}.
\eeq
By comparing with (\ref{limit_beta_0}) and (\ref{rules_c2_1}), one sees that the analysis is analogous to the previous case, by replacing $\A$, $\D$, $\F$ and $\I$ by $\B$, $\E$, $\G$ and $\J$, respectively. 

When the curvaton contribution to the power spectrum is not negligible, significant non-Gaussianity, while satisfying the isocurvature bound, is obtained when $|\G|\ll |\B|\ll 1$. This constraint is  satisfied if one assumes  $\fcc=1-\O(\veps)$, which means that the second curvaton is created mainly by the decay of the first, while $\xrs=1-\O(\veps)$, $\xcc\lesssim\O(\veps)$ and $\fcs=1-\O(\veps)$. 
Other possibilities exist but require some fine-tuning between the parameters,  in analogy with the previous analysis in the case $\betaL \ll 1$.

\subsubsection{Intermediate values of  $\betaL$}
In this case, one must satisfy simultaneously the constraints $|\F|\ll |\A|$ and $|\G|\ll |\B|$, due to the isocurvature bound. 
The relative strength of the different $\fNLt$ coefficients cannot be expressed in such a simple form as in (\ref{rules_c2_1}), but it will be determined again by the ratios $\F/\A$, $\G/\B$, $\I/\D$ and $\J/\E$.

In order to get also a significant non-Gaussianity, we  look for parameter values such that 
\beq
\A, \B \sim \O(\veps), \qquad \F, \G \lesssim \O(\alpha\,  \veps).
\eeq
These constraints can  be satisfied by   fine-tuning  the parameters. Solving $\F\simeq 0$ and $\G\simeq 0$ for the two parameters $\fcf$ and $\fcs$ yields 
\beq 
\fcf\simeq \frac{(\xrf-\xcf)(1-\fcc)-\xcc}{1-\fcc-\xcc} \,, \qquad 
\fcs\simeq \xrs-\xcs+ \frac{1-\xrs}{1-\fcc}\, \xcc\, .
\eeq
The observational constraint on the isocurvature power spectrum is satisfied if  these two fine-tuning relations hold simultaneously, at the level $\O(\alpha \,\varepsilon)$.  Using these relations, one finds
interesting non-Gaussianity for the following set of parameters: $\xrf=\O(\veps)$, $\xrs=\O(\veps)$, $\xcc=\O(\alpha\, \veps)$,
$\fcf=\xrf-\xcf+\O(\alpha\, \veps)$, $\fcs=\xrs+\O(\alpha\, \veps)$, with negligible values for $\xcs$. In this scenario, both curvatons are subdominant at their decay and the fraction of produced dark matter is fine-tuned.

\section{Conclusions}

In this work, we have introduced a systematic treatment in order to compute the evolution of linear and non-linear cosmological perturbations in a cosmological transition due to the decay of some particle species. Our main results can be summarized as follows.

At  the linear level, the evolution   of all individual curvature  perturbations can be expressed in terms of a transfer matrix, whose coefficients depend on background quantities, such as the relative abundances of the fluids at the decay, their equation of state parameters and the relative decay branching ratios
[see Eqs (\ref{decay_linear}-\ref{f_A})]. At the non-linear level, the post-decay curvature perturbations can also be given in terms of the pre-decay perturbations quite generally, and we have presented explicitly these relations at second order [see Eqs (\ref{zeta_2ndorder}-\ref{U_ABC})]. 
We have then applied our general formalism to two specific examples.

The first example is the mixed curvaton-inflaton scenario in which we allow the dark matter to be created both before {\it and} during the curvaton decay. 
 We find, in particular, the remarkable result that it is possible to obtain {\it isocurvature dominated} non-Gaussianities with, as required by the  CMB measurements,  an adiabatic dominated power spectrum.

In the second example, we have studied scenarios with several curvaton-like fields and obtained  results that generalize  previous works on two-curvaton scenarios by taking into account the various decay products of the curvatons. We have explored  the parameter space to see whether  it is possible to find significant non-Gaussianity while satisfying the isocurvature bound in the power spectrum. We have found that several such regions exist, but often at the price of a fine-tuning between the parameters.

In the presence of isocurvature modes, which can be correlated with the adiabatic modes, non-Gaussianity of the local type is much richer than in the purely adiabatic case and we have shown that the angular bispectrum is the sum of six different contributions. As a consequence, in addition to the traditional $f_{NL}$ (adiabatic) parameter, we have identified five new non-linear parameters that must be taken into account: one purely isocurvature parameter and four correlated parameters. We have computed these parameters in the two models we have investigated.

Beyond the two examples considered in this work, our formalism can be used as a general toolbox to  study systematically the cosmological constraints, arising from linear perturbations and from non-Gaussianities, for particle physics models and their associated cosmological scenarios.

\vspace{1cm}

{\bf \noindent Acknowledgements}
The work of D.L.  is partially supported by the ANR  (Agence Nationale de la
Recherche) grant "STR-COSMO" (ANR-09-BLAN-0157).
The work of A. L. is partially supported by the International Doctorate
on AstroParticle Physics (IDAPP) program.

\smallskip

\appendix
\def\n{{\bf \hat n}}

\section{Angular bispectrum}
We consider several observable quantities $X^I$, like $\zeta$ and $S_c$, which depend on  ``primordial'' scalar fields $\phi^a$, whose perturbations are generated during inflation. Up to second order, one can formally write
\beq
\label{X_I}
X^I= N^I_a \phi^a+\frac12 N^{I}_{ab} \phi^a \phi^b + \dots
\eeq
We assume that the $\phi^a$ are Gaussian random fields, with the two-point correlation functions 
\beq
\label{P_ab}
 \langle \phi^a (\vec{k}) \phi^b (\vec{k}')\rangle=
 (2\pi)^3 \, P^{ab}(k) \, \delta(\vec k + \vec k') \,.
 \eeq
 
 We then define the bispectra of the $X^I$ by 
 \beq
\langle X^{I}_{\k_1} X^J_{\k_2} X^{K}_{\k_3} \rangle = (2 \pi)^3 \delta (\Sigma_i \vec k_i) B^{IJK}(k_1, k_2, k_3)\,.
 \eeq
 Substituting the decomposition (\ref{X_I}) into the left hand side, and using (\ref{P_ab}), 
 one finds 
 \begin{eqnarray}
 B^{IJK}(k_1, k_2, k_3)&=&N_a^I N_b^J N_{cd}^K P^{ac}(k_1) P^{bd}(k_2)+N_a^IN_{bc}^JN_d^KP^{ab}(k_1) P^{cd}(k_3)
 \cr
 && 
 +N^I_{ab}N^J_cN^K_d P^{ac}(k_2)P^{bd}(k_3).
 \end{eqnarray}
 
As shown in  \cite{Komatsu:2001rj}, the angular bispectrum can be expressed in terms of the "reduced bispectrum"  $b_{l_1l_2 l_3}$, according to the expression
\beq
\label{angular_3pt}
\langle a_{l_1 m_1} a_{l_2 m_2} a_{l_3 m_3}\rangle ={\cal G}^{m_1m_2m_3}_{l_1l_2l_3}b_{l_1l_2l_3},
\eeq
where 
\beq
{\cal G}^{m_1m_2m_3}_{l_1l_2l_3}\equiv \int d^2\n\,  Y_{l_1m_1}(\n)\,  Y_{l_2m_2}(\n)\,  Y_{l_3m_3}(\n)
\eeq
is the Gaunt integral. 

The next step is to express the reduced bispectrum in terms of the generalized bispectra $B^{IJK}$, using the fact that the observable quantity is related to the initial perturbations $X^I$ via a transfer function $g^I_l(k)$, so that 
\beq
a_{lm}=4\pi (-i)^l \int \frac{d^3\k}{(2\pi)^3} \left(\sum_I X^I(\k) g^I_l(k)\right) Y^*_{lm}(\hat\k).
\eeq
Substituting the above expression in the left hand side of (\ref{angular_3pt}), one finally obtains
\beq
\label{reduced_bispectrum_general}
b_{l_1l_2 l_3}=3\sum_{I,J,K} N^I_{ab} N^J_c N^K_{d} \int_0^\infty r^2 dr \tilde\beta^I_{(l_1}(r)\beta^{J, ac}_{l_2}(r)\beta^{K, bd}_{l_3)}(r),
\eeq
with  
\beq
\tilde\beta^I_{l}(r)\equiv \frac{2}{\pi} \int k^2 dk  j_l(kr) g^I_{l}(k) , \qquad 
\beta^{I,ab}_{l}(r)\equiv \frac{2}{\pi}\int k^2 dk j_l(kr) g^I_{l}(k) P^{ab}(k)\, .
\eeq
Note that the ``reduced'' bispectrum is symmetric with respect to permutations of the indices $l_1$, $l_2$ and $l_3$ (we use the standard notation: 
$(l_1 l_2 l_3)\equiv [l_1l_2l_3+ 5\,  {\rm perms}]/3!$).

In the simplest case, one considers only the adiabatic mode,  $\zeta$ (or the gravitational potential $\Phi$), which is assumed to depend on a single  ``primordial'' Gaussian field. In this case, where both the indices $I$ and $a$ take a single value, one recovers immediately the familiar result of \cite{Komatsu:2001rj}.
Our general expression also  includes the particular situation considered in \cite{Hikage:2008sk}, where $\zeta=\phi+(3/5)f_{NL}\phi^2$ and $S=\eta+f_{NL}^{\rm iso}\eta^2$, $\phi$ and $\eta$ being Gaussian variables.


\end{document}